\pdfoutput=1

\documentclass[11pt]{article}

\usepackage[preprint]{acl}

\usepackage{times}
\usepackage{latexsym}
\usepackage{ulem}
\usepackage[T1]{fontenc}

\usepackage[utf8]{inputenc}

\usepackage{microtype}
\usepackage{wrapfig}
\usepackage{inconsolata}

\usepackage{graphicx}
\usepackage{caption}
\usepackage{booktabs}
\usepackage{multirow}
\usepackage{alltt}
\usepackage{bbding}
\usepackage{enumitem}
\usepackage{amsthm,amsmath,amssymb}
\usepackage{mathrsfs}
\usepackage{tablefootnote}
\usepackage[most]{tcolorbox}
\tcbset{
  aibox/.style={
    width=474.18663pt,
    top=15pt,
    colback=white,
    colframe=black,
    colbacktitle=black,
    fonttitle=\bfseries,
    center,
    attach boxed title to top left={yshift=-0.1in,xshift=0.15in},
    boxed title style={boxrule=0pt,colframe=white,},
  }
}
\newtcolorbox{AIbox}[2][]{aibox,title=#2,#1}
\newtcolorbox{response}[1][]{
  width=210pt,
  colback=gray!5,
  colframe=black,
  fonttitle=\bfseries,
  coltitle=black,
  }
\usepackage{soul}
\title{KnowledgeSG: Privacy-Preserving Synthetic Text Generation with Knowledge Distillation from Server}

\author{
 \textbf{Wenhao Wang\textsuperscript{1,3,4}},
 \textbf{Xiaoyu Liang\textsuperscript{1}},
 \textbf{Rui Ye\textsuperscript{2,4}},
 \textbf{Jingyi Chai\textsuperscript{2,4}}, \\
 \textbf{Siheng Chen\textsuperscript{2,3,4 *}},
 \textbf{Yanfeng Wang\textsuperscript{2,3 *}},
\\
\textsuperscript{1}Zhejiang University, 
\textsuperscript{2}Shanghai Jiao Tong University, \\
\textsuperscript{3}Shanghai AI Laboratory, \\
\textsuperscript{4}Multi-Agent Governance \& Intelligence Crew (MAGIC)
\\
\small{12321254@zju.edu.cn}
}

\begin{document}
\maketitle
\begingroup
\renewcommand\thefootnote{$*$}
\footnotetext{Corresponding author.}
\endgroup

\begin{abstract}
The success of large language models (LLMs) facilitate many parties to fine-tune LLMs on their own private data. However, this practice raises privacy concerns due to the memorization of LLMs. 
Existing solutions, such as utilizing synthetic data for substitution, struggle to simultaneously improve performance and preserve privacy.
They either rely on a local model for generation, resulting in a performance decline, or take advantage of APIs, directly exposing the data to API servers. 
To address this issue, we propose \textit{KnowledgeSG}, a novel client-server framework which enhances synthetic data quality and improves model performance while ensuring privacy. 
We achieve this by learning local knowledge from the private data with differential privacy (DP) and distilling professional knowledge from the server. Additionally, inspired by federated learning, we transmit models rather than data between the client and server to prevent privacy leakage.
Extensive experiments in medical and financial domains demonstrate the effectiveness of \textit{KnowledgeSG}. Our code is now publicly available at \href{https://github.com/wwh0411/KnowledgeSG}{https://github.com/wwh0411/KnowledgeSG}. 

\end{abstract}

\section{Introduction}
\label{sec:intro}

The world has witnessed the tremendous success of large language models (LLMs) across a variety of tasks~\cite{llama2,openai2023gpt4}.
Such success has attracted numerous parties to fine-tune their customized LLMs by leveraging their local private data~\cite{wu2023bloomberggpt,xue2023db,lawgpt,singhal2023large}.
Nonetheless, training such LLMs on private data could cause significant privacy concerns, since LLMs are shown to memorize sensitive information from the training data~\cite{carlini2021extracting, lukasAnalyzingLeakagePersonally2023}.

To address this privacy issue, a series of methods have been proposed to circumvent the direct usage of private data by 
using synthetic data for substitution \cite{xieDifferentiallyPrivateSynthetic2024,yueSyntheticTextGeneration2023, liSyntheticDataAlmost2024}. Specifically, some methods use Application Programming Interface (APIs) to generate diverse instructions, directly exposing private data to the API server \cite{wang2022self}. While others rely solely on a local base model, which leads to a quality degradation in synthetic data and eventually lower model performance  \cite{kurakinHarnessingLargelanguageModels2024}.
Therefore, existing methods suffer from the trade-off between privacy risk and model performance.

\begin{figure}[t]
    \centering
    \includegraphics[width=1\columnwidth]
    {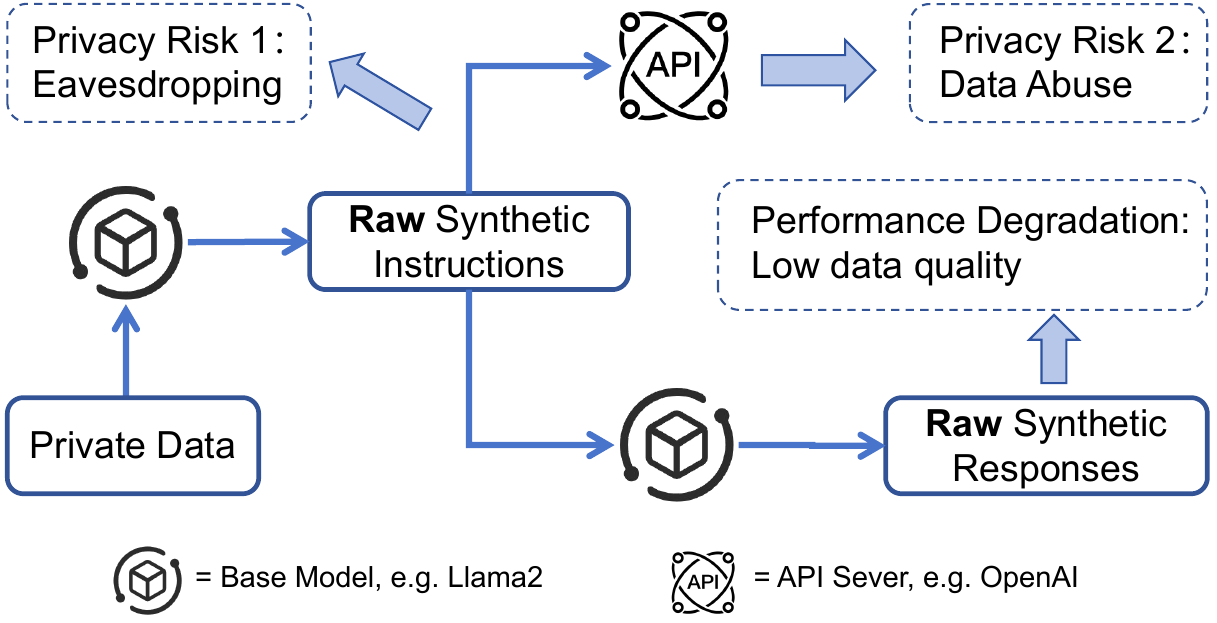}
    \caption{The dilemma of current synthetic data methods. API-based methods involve more privacy risks while methods based on local models face performance degradation due to lower synthetic data quality.}
    \label{fig:comparison}
    \vspace{-2.5mm}
\end{figure}

In this work, we aim to efficiently enhance synthetic data quality
while maintaining strict privacy protection. To achieve this goal, we propose \textit{KnowledgeSG} (\textbf{Knowledge}-based \textbf{S}ynthetic data \textbf{G}eneration), a novel client-server framework which leverages a professional server to assist the local client in data generation under theoretical privacy guarantee. 
Our framework compensates the quality gap between synthetic and original data observed in previous works \cite{jordon2022synthetic, arnold2021really} by efficiently distilling knowledge from the professional model deployed on the server, rather than relying merely on the local model.
Additionally, unlike API-based methods, we draw inspiration from federated learning \cite{fedavg} by transmitting model weights instead of data for knowledge exchange, thereby improving privacy protection.

Specifically, on the client side, we fine-tune the local model with differentially privacy (DP) to learn local knowledge from private data within a privacy budget. 
For convenient and secure communication between the client and server, we transmit only the LoRA \cite{hu2021lora} adapter of the DP-finetuned model instead of directly transmitting private data. 
On the server side, raw synthetic instructions are first generated using the uploaded local model. These instructions are then judged by the professional model for quality filtration in an efficient manner \cite{jiang2023lion}. Once filtered, the top instructions are fed directly into the professional model to generate accurate responses, bypassing the need to generate potentially incorrect responses from the local model.
Finally, the DP-finetuned local model is further optimized by fine-tuning it with the top instructions and corresponding responses to boost its performance. Upon completion, the optimized model is transmitted back to the client, concluding the entire process.

We conduct a series of experiments on two privacy-sensitive domains: medicine and finance. The results prove the effectiveness of our proposed framework on both privacy and performance benchmarks.
It is worth mentioning that our method gains a relative improvement of $120.39\%$ than Non-Private training measured by medical free-form evaluation, even surpassing AlpaCare \cite{zhang2023alpacareinstructiontuned}, the professional model we deploy. 
To conclude, our main contributions are:
\begin{enumerate}[nosep]
    \item We propose a novel privacy-preserving client-server framework called \textit{KnowledgeSG}, which enhances synthetic data quality by leveraging server-side knowledge distillation to assist the client in data generation.
    \item We propose a novel server-side synthetic data generation method that employs a professional model to distill knowledge by providing both judgments and corrections for the raw synthetic data.
    \item Extensive experiments validate the effectiveness of our proposed framework.
\end{enumerate}

\begin{figure*}[t]
\vspace{-2mm}
    \centering
    \includegraphics[width=0.9\textwidth]
    {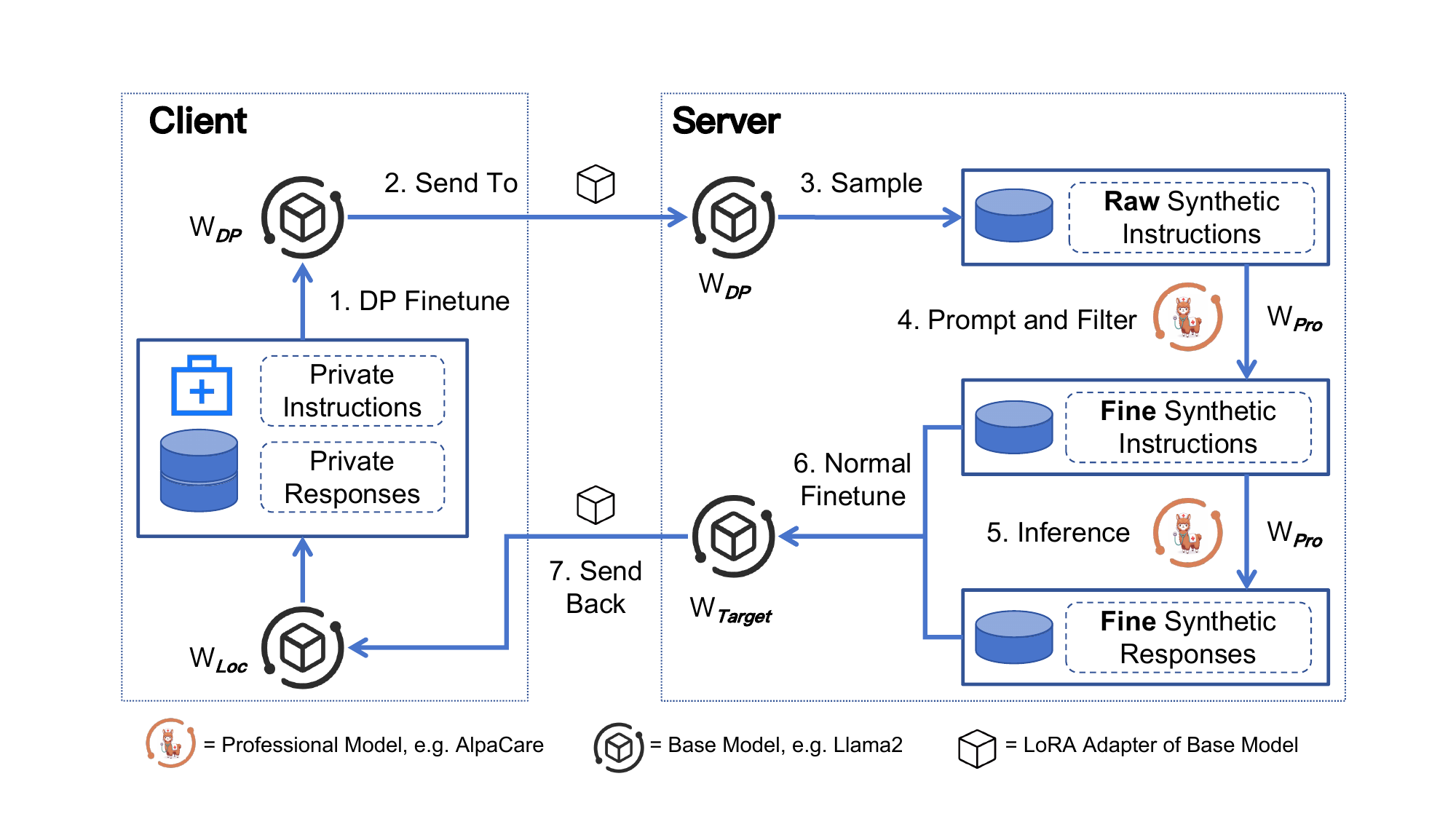}
    \caption{Overview of \textit{KnowledgeSG}'s system architecture. $\mathbb{W}_{Loc}$: the local base model; $\mathbb{W}_{DP}$: DP-finetuned $\mathbb{W}_{Loc}$; $\mathbb{W}_{Target}$: the final target model; $\mathbb{W}_{Pro}$: the professional model. From left to right, $\mathbb{W}_{Loc}$ learns knowledge from private data on the client side and acquires knowledge distillation from $\mathbb{W}_{Pro}$ on the server side.}
    \label{fig:system_architecture}
    \vspace{-2.5mm}
\end{figure*}

\section{Related Work}
\label{sec:related_work}
\subsection{Privacy Concerns with Fine-tuning on Private Data}
Fine-tuning large language models is crucial for enhancing their instruction following ability and improving performance on certain downstream tasks \cite{DatabricksBlog2023DollyV2, wang2023far, pmlr-v202-jang23a}. 
In order to deliver a satisfactory user experience \cite{zhao2024wildchat} or achieve professional-level expertise \cite{codealpaca, xu2023wizardlm}, it is inevitable to fine-tune LLMs on user-related private data or proprietary data owned by institutions.
However, recent studies \cite{kandpalUserInferenceAttacks2023, carlini2021extracting} have experimentally demonstrated that LLMs can memorize their training datasets, leaving possibilities of leaking private information through either simple prompts \cite{carlini2021extracting, nasrScalableExtractionTraining2023} or delicately designed attacks \cite{lukasAnalyzingLeakagePersonally2023, guptaRecoveringPrivateText2022}.

Continuing to improve the quality and coverage of fine-tuned large language models necessitates the development of alternative approaches to utilizing private data without memorizing it.
To mitigate this issue, two mainstream solutions have emerged. The first involves fine-tuning LLMs with differential privacy techniques \cite{Abadi_2016, yuDifferentiallyPrivateFinetuning2022}, while the second focuses on substituting original private data with high-fidelity synthetic ones for fine-tuning \cite{yueSyntheticTextGeneration2023, xieDifferentiallyPrivateSynthetic2024}.

\subsection{Synthetic Text Generation}
Two widely adopted approaches for generating private synthetic text in practice are In-Context Learning (ICL) \cite{in-context, chen2024retrievalstyleincontextlearningfewshot, ye2024leveraging} and Self-Instruction \cite{wang2022self}. 
Largely relying on prompt design and the base model's comprehension, they suffer from either low data fidelity yielded by the base model, or privacy concerns requesting API servers. What makes it worse, with private data included directly in prompts, these methods pose an additional risk of revealing sensitive information.

Recently, researchers have recognized the feasibility and effectiveness of the DP generator method \cite{yuPrivacyPreservingInstructionsAligning2024, yueSyntheticTextGeneration2023, kurakinHarnessingLargelanguageModels2024}. This approach first trains an LLM on private data with DP, and then repeatedly samples the DP-finetuned model to generate synthetic text sequences.
Although proved to gain improvements in distribution similarity, previous works primarily concentrate on generating diverse synthetic instructions. They ignore or skip the practical scenarios where responses are equally crucial for instruction tuning of LLMs. 
Moreover, current DP generator methods only focus on general knowledge, leading to significantly poorer performance in domain-specific scenarios such as finance and medicine where privacy draws considerable attention.
Therefore, \textit{KnowledgeSG} intends to improve the quality of both synthetic instructions and responses by distilling the professional model, especially on domain-specific tasks.

\section{Method}
\label{sec:method}
\subsection{Problem Setup}
\label{sec:problem_setup}
Let $\mathbb{D}_{Pri}$ represent the private dataset possessed by the client, which contains privacy from patients. $\mathbb{W}_{Loc}$ is the local base model pre-trained on general data that needs to acquire medical knowledge from $\mathbb{D}_{Pri}$. $\mathbb{W}_{Pro}$ refers to the professional model hosted by the server which is relatively larger than $\mathbb{W}_{Loc}$ and is assumed to have extensive knowledge of the medical domain.
To formalize our problem setup, we assume that $\mathbb{D}_{Pri}$ used for instruction tuning consists of two components: \textit{Instruction} and \textit{Response}, both of which contain Personal Identifiable Information (PII), e.g. patients' names. Therefore, $\mathbb{D}_{Pri}$ can not be directly transmitted over networks due to privacy concerns. We present a detailed definition of PII in Appendix \ref{sec:pii}. 

Our ultimate objective is to generate a synthetic dataset $\mathbb{D}_{Syn}$ that maintains high data quality while containing no trace of PIIs. This allows us to fine-tune $\mathbb{W}_{Loc}$ on $\mathbb{D}_{Syn}$ to facilitate improvements in privacy-performance trade-off. 

\subsection{System Overview}
We introduce a novel client-server framework called \textit{KnowledgeSG} (\textbf{Knowledge}-based \textbf{S}ynthetic data \textbf{G}eneration), which aims to improve synthetic data quality and further promote model performance without violating privacy.

We attribute the quality gap between synthetic data and original private data to the comprehension deficiency of the local model $\mathbb{W}_{Loc}$ used for generation.
Due to privacy concern, previous works place all generation on the client side without involving the server. 
To compensate for the aforementioned comprehension deficiency, we further extend previous setting into a client-server framework to leverage the knowledge from the server-side professional model $\mathbb{W}_{Pro}$. We give further elaboration of the quality gap in Appendix \ref{sec:gap}.

The client-server framework of \textit{KnowledgeSG} involves learning local knowledge from private data on the client side and acquiring knowledge distillation from the professional model on the server side. We also design a convenient transmitting unit to mitigate potential eavesdropping.
In this way, we manage to achieve superior performance results while preventing memorization or leakage of the private dataset $\mathbb{D}_{Pri}$.

\subsection{Client Side}
\label{sec:client_side}
On the client side, our framework is primarily designed to extract knowledge from the private data $\mathbb{D}_{Pri}$ without memorization and subordinately designed to be lightweight. 

\paragraph{DP-based Local Learning.}
Due to its direct access to $\mathbb{D}_{Pri}$, the client side must comply with strict privacy constraint while still enabling effective knowledge learning from the private dataset $\mathbb{D}_{Pri}$.
To achieve this primary goal, we adopt Differentially Private SGD (DP-SGD) \cite{Abadi_2016}. 

DP-SGD is a privacy-preserving optimization algorithm that improves upon traditional Stochastic Gradient Descend (SGD) by adding noise to the gradients during training. This noise ensures that the inclusion or exclusion of any individual data sample has a minimal impact on the resulting fine-tuned model $\mathbb{W}_{DP}$, offering strong privacy guarantees.
We follow the first step of previous works \cite{yuDifferentiallyPrivateFinetuning2022, kurakinHarnessingLargelanguageModels2024, yueSyntheticTextGeneration2023} and adopt DP-SGD as our local training approach.
The local base model $\mathbb{W}_{Loc}$ pre-trained on general corpora, is fine-tuned through DP-SGD, i.e. DP-finetuned on $\mathbb{D}_{Pri}$ to gain local knowledge under a privacy budget $(\epsilon,\delta)-DP$. 
This budget theoretically guarantees the process of DP-finetuning without any leakage of private information, providing the basis for us to transmit the fine-tuned model $\mathbb{W}_{DP}$ to the server later.

\paragraph{LoRA Adaptation.}
The second characteristic of the client side in \textit{Knowledge} is lightweight, since we do not expect the client to have substantial hardware resources compared to the server. 
Therefore, we minimize the workload on the client by shifting the resource-intensive data generation process to the server.

Besides, we apply Low-Rank Adaptation (LoRA) \cite{hu2021lora} using the implementation of \citet{dp-transformers}, as our training approach. 
LoRA is an efficient fine-tuning technique for large language models. It reduces the number of trainable parameters by introducing low-rank decomposition into the weight matrices of the model, allowing for faster and more resource-efficient adaptation to new tasks. 

Even when considered relatively "small", the full size of the base model such as Llama2-7B, still occupies a significant amount of storage. 
The resulting inconvenience for transmitting the full model weights of $\mathbb{W}_{DP}$ is plain to see. 
In contrast, LoRA adaptation significantly reduces the transmission burden by allowing us to send only the LoRA adapter $\mathbb{A}_{DP}$, resulting in a far more manageable model size.
Detailed comparison of model sizes is shown in Table \ref{tab:size}. 

\begin{table}[t]
\setlength\tabcolsep{4pt}
\centering
\small
\setlength\tabcolsep{3.3pt}
\begin{tabular}{ccc}
\toprule 
Model Type & Params & Size \\
\midrule
Base Model & 6738 M & 26 GB \\
LoRA Adapter & 4.2 M & 33 MB \\
\bottomrule
\end{tabular}
\caption{The parameter numbers and model sizes for Llama2-7B with \& without LoRA rank of $16$.}
\label{tab:size}
\vspace{-3mm}
\end{table}

\subsection{Server Side}
The server side of \textit{KnowledgeSG} is designed to improve data quality beyond what can be achieved by relying solely on the client. It operates through three stages: raw synthetic data generation, refinement of raw synthetic data and normal fine-tuning of local model. 

\paragraph{Synthetic Instructions Generation.} 
The first step on the server side is to recover the full model $\mathbb{W}_{DP}$ from $\mathbb{A}_{DP}$, assuming the server has the same base model $\mathbb{W}_{Loc}$ as the client prior to communication.
Afterward, we prompt the DP-finetuned model $\mathbb{W}_{DP}$, which has knowledge of the private data $\mathbb{D}_{Pri}$, to generate raw synthetic instructions. 

The post-processing property of DP \cite{Dwork2014TheAF} ensures that once the model $\mathbb{W}_{Loc}$ has been fine-tuned with DP, sampling from the fine-tuned model $\mathbb{W}_{DP}$ incurs no extra privacy loss. As a result, when the LoRA adapter $\mathbb{A}_{DP}$ is uploaded to the server, it can generate synthetic data without exceeding the privacy budget $(\epsilon,\delta)-DP$. 

\paragraph{Synthetic Instruction Filtration.}
During the second stage, to realize optimal results, we apply two compatible filtration methods distinguished by whether assistance from the professional model $\mathbb{W}_{Pro}$ is required.

Filtration without $\mathbb{W}_{Pro}$ uses similarity de-duplication via the BLEU score \cite{Papineni02bleu:a}.
Bilingual Evaluation Understudy (BLEU) is a widely used automated evaluation metric for measuring the similarity between machine translation outputs and reference translations to assess translation quality. We adopt it to determine if an synthetic instruction is too similar to any example from the private dataset $\mathbb{D}_{Pri}$ to raise possibilities of leaking privacy. This method is much faster compared with the other model-based method.

For the filtration method involving $\mathbb{W}_{Pro}$, we prompt the raw instructions into $\mathbb{W}_{Pro}$ for judgements. If the instruction is domain-specific, $\mathbb{W}_{Pro}$ assesses whether it is relevant to its domain. 
If it is domain-specific, $\mathbb{W}_{Pro}$ judges an instructions based on whether this instruction is related to its domain. The detailed prompt we use is provided in Appendix \ref{sec:template}.
 
\paragraph{Efficient Knowledge Distillation.}
Without the need to derive loss from $\mathbb{W}_{Pro}$ \cite{flemings2024differentially}, we use a convenient method of knowledge distillation by feeding top instructions into $\mathbb{W}_{Pro}$ to generate preferable responses corresponding to these instructions  after filtration \cite{xu2023wizardlm, wang2022self, jiang2023lion}.
This step is crucial as the knowledge is embedded in these responses which are subsequently distilled into the local model $\mathbb{W}_{DP}$ through fine-tuning. 

Finally, we use the generated instructions and responses sorted by the IFD score \cite{Li2023FromQT} to normally (non-DP) fine-tune $\mathbb{W}_{DP}$ and obtain the desired model $\mathbb{W}_{Target}$. 
Further details and results regarding the IFD score are presented in Section \ref{sec:data_quality}.
At this stage, DP-finetuning is not needed, as we assume the refined synthetic data contains no sensitive information.

\subsection{Communication between Client \& Server}
\paragraph{Federated Model Transmission.}
Although synthetic data is supposed to contain no privacy, i.e. PIIs, two non-negligible concerns remain: 
(1) The size of the data prepared for fine-tuning are relatively larger than that of the LoRA adapter $\mathbb{A}_{DP}$. 
(2) Leakage of synthetic data can potentially reveal approximate data distribution or other sensitive information.

Therefore, inspired by federated fine-tuning of language models \cite{wei2020federated, yeOpenFedLLMTrainingLarge2024}, we propose to apply transmitting the fine-tuned version of model into our new setting which only has one client and one server, rather than directly transmitting data. 

\paragraph{Proposed Transmitting Unit.}
Moreover, to reduce the potential risk of eavesdropping, i.e. an unauthorized party intercepts and steals the transmitted model, we introduce an efficient transmitting unit. Note that this unit is compatible and optional if the client using \textit{KnowledgeSG} has no concerns about eavesdropping.

We start by sampling a small amount of data from public datasets, e.g. Alpaca \cite{alpaca}, as the seed dataset $\mathbb{D}_{Seed}$, which is agreed and shared by the client and server at the beginning. Then we fine-tune the original base model $\mathbb{W}_{Loc}$ on $\mathbb{D}_{Seed}$ to create a full adaption of model weights and replace original $\mathbb{W}_{Loc}$ with the new model $\mathbb{W}^{'}_{Loc}$. The local learning process described in Section \ref{sec:client_side} is based on $\mathbb{W}^{'}_{Loc}$ afterwards. 
In this way, we make sure that, even if an adversarial eavesdropper intercepts the LoRA adapter $\mathbb{A}_{DP}$, he cannot recover our entire model with the old version of base model $\mathbb{W}_{Loc}$ instead of $\mathbb{W}^{'}_{Loc}$. 

\section{Experiments}
\label{experiment}
\subsection{Basic Setups}
\label{sec:basic_setups}
\paragraph{Models and Datasets.}
If not otherwise mentioned, our base model is pre-trained Llama2-7B \cite{llama2}.
We choose FinGPT \cite{fingpt_github} and AlpaCare \cite{zhang2023alpacareinstructiontuned} as our professional models for financial and medical domains respectively.
The dataset sample is kept to $500$ for any comparison except the ablation study in Section \ref{sec:size}.
We use the name substitution technique in Appendix \ref{sec:name_substitution} to pre-process datasets, preventing inaccurate evaluation on privacy.

\paragraph{Baselines.}
\label{baseline}
Our baselines comprise one None-Private approach, one private approach with DP-SGD \cite{Abadi_2016}, and six private approaches using synthetic data generation, i.e. ICL \cite{in-context}, Self-Instruct \cite{wang2022self}, Self-Instruct-ICL, DP-Gene \cite{kurakinHarnessingLargelanguageModels2024}, DP-Instruct \cite{yuPrivacyPreservingInstructionsAligning2024} and DP-Instruct-ICL. 
The detailed comparison of baselines is shown in Table \ref{tab:baseline} in Appendix \ref{sec:appendix_baseline}.

\subsection{Privacy Evaluation}
\label{sec:privacy_evaluation}
\definecolor{LemonChiffon}{RGB}{255, 240, 205}
\definecolor{lavender}{RGB}{230, 230, 250}
\begin{table*}[ht]
\vspace{-3mm}
\small
\centering
\begin{tabular}{l|r l@{\hspace{4.6\tabcolsep}}r| r l@{\hspace{4.6\tabcolsep}}r}
    \toprule
    Baselines & \multicolumn{2}{c}{\textcolor[RGB]{172, 172, 234}{Medical}} & Inc & \multicolumn{2}{c}{\textcolor[RGB]{241, 205, 122}{Financial}} & Inc \\
    \midrule
    Random & 1.56 & \begin{tikzpicture}\filldraw[lavender] (0,-0.1) rectangle (0.156*2,0.12);\end{tikzpicture} & 0& 1.56 & \begin{tikzpicture}\filldraw[LemonChiffon] (0,-0.1) rectangle (0.156*2,0.12);\end{tikzpicture} & 0 \\
    \midrule
    Non-Private & 97.13 & \begin{tikzpicture}\filldraw[lavender] (0,-0.1) rectangle (1.75*2,0.12);\end{tikzpicture} & 95.57 & 96.23 & \begin{tikzpicture}\filldraw[LemonChiffon] (0,-0.1) rectangle (1.75*2,0.12);\end{tikzpicture} & 94.67 \\
    ICL & 5.47 & \begin{tikzpicture}\filldraw[lavender] (0,-0.1) rectangle (0.547*2,0.12);\end{tikzpicture} & 3.91 & 7.40 & \begin{tikzpicture}\filldraw[LemonChiffon] (0,-0.1) rectangle (0.547*2,0.12);\end{tikzpicture} & 5.84 \\
    Self-Instruct & 1.46 & \begin{tikzpicture}\filldraw[lavender] (0,-0.1) rectangle (0.146*2,0.12);\end{tikzpicture} & -0.10 & 1.89 & \begin{tikzpicture}\filldraw[LemonChiffon] (0,-0.1) rectangle (0.189*2,0.12);\end{tikzpicture} & 0.33 \\
    Self-Instruct-ICL & 3.33 & \begin{tikzpicture}\filldraw[lavender] (0,-0.1) rectangle (0.333*2,0.12);\end{tikzpicture} & 1.77 & 3.77 & \begin{tikzpicture}\filldraw[LemonChiffon] (0,-0.1) rectangle (0.337*2,0.12);\end{tikzpicture} & 1.81 \\
    DP-Gene & 2.26 & \begin{tikzpicture}\filldraw[lavender] (0,-0.1) rectangle (0.226*2,0.12);\end{tikzpicture} & 0.70 & 2.52 & \begin{tikzpicture}\filldraw[LemonChiffon] (0,-0.1) rectangle (0.252*2,0.12);\end{tikzpicture} & 0.96 \\
    DP-Instruct & 1.07 & \begin{tikzpicture}\filldraw[lavender] (0,-0.1) rectangle (0.107*2,0.12);\end{tikzpicture} & -0.49 & 3.14 & \begin{tikzpicture}\filldraw[LemonChiffon] (0,-0.1) rectangle (0.314*2,0.12);\end{tikzpicture} & 1.58 \\
    DP-Instruct-ICL & 3.60 & \begin{tikzpicture}\filldraw[lavender] (0,-0.1) rectangle (0.360*2,0.12);\end{tikzpicture} & 2.04 & 5.03 & \begin{tikzpicture}\filldraw[LemonChiffon] (0,-0.1) rectangle (0.503*2,0.12);\end{tikzpicture} & 3.47 \\
    KnowledgeSG & 0.87 & \begin{tikzpicture}\filldraw[lavender] (0,-0.1) rectangle (0.087*2,0.12);\end{tikzpicture} & -0.69 & 1.89 & \begin{tikzpicture}\filldraw[LemonChiffon] (0,-0.1) rectangle (0.189*2,0.12);\end{tikzpicture} & 0.33 \\
    
    \bottomrule
\end{tabular}
\caption{Reconstruction rate comparison between different baselines on the medical and financial domains. \textit{Inc} represents the increase of reconstruction rate between certain baseline and random guessing. Higher reconstruction rate indicates more memorization of the private data. Results in both domains demonstrate that synthetic data methods, including \textit{KnowledgeSG}, achieve significantly better privacy protection than non-private methods.}

\label{tab:privacy_baseline}
\vspace{-2.5mm}
\end{table*}
\paragraph{Setups.}
We study the privacy leakage of LLM by measuring the reconstruction rates following \citet{lukasAnalyzingLeakagePersonally2023}\footnote{\href{https://github.com/microsoft/analysing_pii_leakage}{https://github.com/microsoft/analysing\_pii\_leakage}}. In this approach, the attacker is given a sentence with multiple masked pieces of PII and asked to reconstruct the target PII from given candidates. 
The reconstruction rate is then calculated as the success ratio over attempt times.

In practice, for each sample in our training dataset, we mask all individual names and randomly choose one as the target. Then we use the PII reconstruction attack \cite{lukasAnalyzingLeakagePersonally2023} to predict the targeted individual name from a list of candidates and report the average prediction accuracy. 
Concretely, each time we sample $64$ names as candidates from our datasets, making sure one of them is correct, and decode from the model using top-k sampling with k set to $40$.
We employ Flair\footnote{\href{https://github.com/flairNLP/flair}{https://github.com/flairNLP/flair}} models \cite{akbik2018coling} to tag individual names in the datasets. 

\paragraph{Results.}

From Table \ref{tab:privacy_baseline}, we can see that:
(1) Using synthetic data instead of original data successfully reduces the PII reconstruction rate by a tremendous margin, demonstrating superior privacy protection over Non-Private method. 
(2) Differentially private training can preserve data privacy to a great content, but is still not on par with synthetic data approaches.
(3) The privacy protection capabilities of different baselines exploiting synthetic data are closely aligned, with \textit{KnowledgeSG} ranking first and ICL lagging behind, which validates the effectiveness of our method. This is reasonable in that ICL-related methods require few-shot examples from the original dataset to generate responses, thus introducing greater privacy risks.

\subsection{Financial Benchmarks}
\label{financial}
\paragraph{Setups.}
We use the financial sentiment analysis dataset\footnote{\href{https://huggingface.co/datasets/FinGPT/fingpt-sentiment-train}{https://huggingface.co/datasets/FinGPT/fingpt-sentiment-train}\label{fingpt}} as the training dataset~\cite{yang2023fingpt}.
During the evaluation, we employ the code from \citet{yang2023fingpt}\footnote{\href{https://github.com/AI4Finance-Foundation/FinGPT}{https://github.com/AI4Finance-Foundation/FinGPT}} and consider four financial sentiment analysis benchmarks, including FPB \cite{Malo2014GoodDO}, FIQA-SA \cite{Maia2018WWW18OC}, TFNS \cite{tfns2022}, and NWGI \cite{fingpt_github}, where both accuracy and F1 score are measured.
Besides, we also report the performance of GPT-3.5 \cite{ouyang2022training} and GPT-4 \cite{openai2023gpt4} for reference.
Since NWGI cannot be measured using GPT-3.5/4, we report the average metric of the first three and four evaluation datasets for an overall comparison.

\paragraph{Results.}
Table \ref{tab:financial_benchmark} demonstrates the results of our method and six other baselines using synthetic data generation on financial benchmarks. From the table, we can conclude that:
(1) \textit{KnowledgeSG} outperforms all other baselines on average, even better than using original private data, proving the effectiveness of knowledge distillation from professional model through our framework, not to mention our privacy-preserving nature.
(2) For the FiQA-SA dataset, a large portion of the evaluation sample labels are \textit{Neutral}. Following the evaluation benchmarks \cite{fingpt_github}, we treat responses with no predictions (Positive/Negative/Neutral) as \textit{Neutral}. This situation rarely happens except for pre-trained models that struggle with instruction following. Most of LLaMA2-7B's responses are classified as \textit{Neutral}, thus explaining its unexpectedly strong performance on FiQA-SA.
(3) Ignoring FiQA-SA, some synthetic generation baselines still perform even worse than the pre-trained Llama2 on FPB and TFNS. This phenomenon shows evidence for the quality issue we found for domain-specific data after generation. The \textit{Gap Ratio}, as introduced in Appendix \ref{sec:gap_ratio} is $0.4682$ for FPB and $0.3663$ for TFNS, both below the heuristically drawn datum line of $0.5$.

\begin{table*}[t]
\vspace{-3mm}

\setlength\tabcolsep{6pt}
\centering
\small
\begin{tabular}{l|cccccccccccc}
\toprule
\multirow{2}{*}{Evaluation} & \multicolumn{2}{c}{FPB} & \multicolumn{2}{c}{FiQA-SA} & \multicolumn{2}{c}{TFNS} & \multicolumn{2}{c}{NWGI} & \multicolumn{2}{c}{Avg:3} & \multicolumn{2}{c}{Avg:4} \\
& Acc & F1 & Acc & F1 & Acc & F1 & Acc & F1 & Acc & F1 & Acc & F1\\
\midrule
GPT-3.5 & 0.781  & 0.781  & 0.662  & 0.730  & 0.731  & 0.736  & - & - & 0.725  & 0.749  & - & -  \\ 
GPT-4 & 0.834  & 0.833  & 0.545  & 0.630  & 0.813  & 0.808  & - & - & 0.731  & 0.757  & - & -  \\ 
Llama2-7B & 0.462  & 0.390  & 0.822  & 0.800  & 0.386  & 0.296  & 0.583  & 0.503  & 0.557  & 0.495  & 0.563  & 0.497   \\ 
FinGPT v3.3 & 0.882  & 0.882  & 0.858  & 0.874  & 0.903  & 0.903  & 0.643  & 0.643  & 0.881  & 0.886  & 0.822  & 0.826   \\ 
\midrule
Non-Private & 0.753  & 0.752  & 0.724  & 0.767  & 0.622  & 0.639  & 0.657  & 0.656  & 0.699  & 0.719  & 0.689  & 0.703   \\ 
ICL & 0.366  & 0.251  & 0.724  & 0.725  & 0.418  & 0.421  & 0.563  & 0.532  & 0.502  & 0.466  & 0.517  & 0.482   \\ 
Self-Instruct & 0.317  & 0.185  & 0.695  & 0.661  & 0.304  & 0.257  & 0.489  & 0.404  & 0.439  & 0.368  & 0.451  & 0.377   \\ 
Self-Instruct-ICL & 0.295  & 0.153  & 0.644  & 0.561  & 0.483  & 0.483  & 0.461  & 0.347  & 0.474  & 0.399  & 0.470  & 0.386   \\ 
DP-Gene & 0.308  & 0.181  & 0.618  & 0.519  & 0.397  & 0.371  & 0.453  & 0.366  & 0.441  & 0.357  & 0.444  & 0.359   \\ 
DP-Instruct & 0.296  & 0.285  & 0.615  & 0.489  & 0.439  & 0.439  & 0.421  & 0.300  & 0.450  & 0.404  & 0.443  & 0.378   \\ 
DP-Instruct-ICL & 0.332  & 0.299  & 0.666  & 0.588  & 0.399  & 0.345  & 0.472  & 0.382  & 0.465  & 0.410  & 0.467  & 0.403  \\ 
KnowledgeSG & \textbf{0.779}  & \textbf{0.775}  & \textbf{0.791}  & \textbf{0.806}  & \textbf{0.782}  & \textbf{0.743}  & \textbf{0.658}  & \textbf{0.658}  & \textbf{0.784}  & \textbf{0.775}  & \textbf{0.752}  & \textbf{0.745}  \\

\bottomrule
\end{tabular}
\caption{Comparison with baselines on the financial benchmarks, where the sentiment analysis dataset from FinGPT~\cite{yang2023fingpt} is used. Four evaluation datasets are considered, including FPB, FIQA-SA, TFNS, and NWGI. We also show results of GPT-3.5/4, Llama2-7B and FinGPT v3.3 for reference. We leverage Llama2-7B as the base model and FinGPT v3.3 as the professional model. The results demonstrate that \textit{KnowledgeSG} outperforms all other baselines and is on par with the performance of GPT3.5/4.}
\label{tab:financial_benchmark}
\vspace{-2.5mm}
\end{table*}

\subsection{Medical Free-Form Evaluation}
\label{sec:medical_freeform}
\paragraph{Setups.}
We utilize the HealthCareMagic-100k dataset\footnote{\href{https://huggingface.co/datasets/lavita/ChatDoctor-HealthCareMagic-100k}{https://huggingface.co/datasets/lavita/ChatDoctor-HealthCareMagic-100k}} \cite{li2023chatdoctor} as our training dataset, since it contains many individual names (e.g. see Fig \ref{fig:icliniq}). This dataset consists of real conversations between patients and doctors collected from the HealthCareMagic website. 

Following \citet{zhang2023alpacareinstructiontuned}, we conduct free-form evaluation by employing GPT-3.5-turbo \cite{zheng2023judging} to serve as a judge. For each instruction in the test dataset, the judge pairwise compares two responses resulting from the target model and THE reference model, respectively.
We employ text-davinci-003, GPT-3.5-turbo, GPT-4 and Claude-2 as reference models.
To avoid positional bias, we evaluate each sample twice with exchanged positions of different responses generated by the test and reference models. We follow \citet{alpaca_eval} to score the models by calculating the win rate. 
Additional experiments on medical benchmarks are attached in Appendix \ref{sec:medical_benchmark}.

\begin{table*}[t]
\centering
\small
\begin{tabular}{l|ccccc}
\toprule
Evaluation & \makebox[0.118\textwidth][c]{Text-davinci-003} & \makebox[0.118\textwidth][c]{GPT-3.5-turbo} & \makebox[0.118\textwidth][c]{GPT-4} & \makebox[0.118\textwidth][c]{Claude-2} & \makebox[0.118\textwidth][c]{Avg} \\
\midrule
AlpaCare \cite{zhang2023alpacareinstructiontuned} & 0.666 & 0.506 & 0.474 & 0.497 & 0.536  \\ 
Llama2-7B & 0.135  & 0.104  & 0.038  & 0.046  & 0.081   \\ 
\midrule
Non-Private & 0.389  & 0.303  & 0.151  & 0.179  & 0.255   \\ 
ICL \cite{in-context} & 0.380  & 0.280  & 0.141  & 0.166  & 0.241   \\ 
Self-Instruct \cite{wang2022self} & 0.208  & 0.152  & 0.054  & 0.054  & 0.117   \\ 
Self-Instruct-ICL & 0.247  & 0.167  & 0.064  & 0.089  & 0.142   \\ 
DP-Gene \cite{kurakinHarnessingLargelanguageModels2024} & 0.307  & 0.235  & 0.097  & 0.121  & 0.190   \\ 
DP-Instruct \cite{yuPrivacyPreservingInstructionsAligning2024} & 0.255  & 0.184  & 0.076  & 0.097  & 0.153   \\ 
DP-Instruct-ICL & 0.382  & 0.295  & 0.187  & 0.199  & 0.266   \\ 
KnowledgeSG & \uline{\textbf{0.776}}  & \uline{\textbf{0.530}} & \textbf{0.457}  & \textbf{0.488}  & \uline{\textbf{0.562}}  \\ 
\bottomrule
\end{tabular}
\caption{Performance results and comparative analysis of free-form instruction evaluation in the medical domain. \textit{KnowledgeSG} outperforms all other baselines and has a relative improvement of $120.39\%$ than Non-Private method. Numbers with underlines represent performance surpassing the professional model AlpaCare \cite{zhang2023alpacareinstructiontuned}.}
\label{tab:medical_alpacare}
\vspace{0mm}
\end{table*}

\paragraph{Results.}
From Table \ref{tab:medical_alpacare} and Table \ref{tab:medical_benchmark}, we can conclude that:
(1) Considering both benchmark and free-form results, \textit{KnowledgeSG} consistently and significantly surpasses all other baselines in the medical domain. 
Particularly in the free-from evaluation, our method outperforms all other synthetic text generation baselines to a large margin, even doubling the performance of the None-private approach using original private data.
(2) DP-based generation methods achieve much higher win rate scores than that of Self-instruction-based methods.
This is expected because DP-based methods require additionally differentially private fine-tuning of the base model on private data. 
(3) The free-form results of \textit{KnowledgeSG} surpassing AlpaCare (underlined in Table \ref{tab:medical_alpacare}) highlight the immense potential of synthetic generation approaches which acquire knowledge distillation from a professional model, inspiring future research to further explore this area. 

\subsection{Data Quality Measurement.}
\label{sec:data_quality}
\paragraph{Embedding Distribution Similarity.}
As shown in \citet{yueSyntheticTextGeneration2023}, the similarity of synthetic data to the original data implicitly indicates its quality.
Unlike typical natural language generation (NLG) tasks such as machine translation, which have ground truth references for evaluation, quantifying the similarity between synthetic and original private samples is non-trivial due to the absence of one-to-one mapping between them. 

To measure the embedding distribution distance between synthetic and original data, we use sentence-transformers\footnote{\href{https://huggingface.co/sentence-transformers}{https://huggingface.co/sentence-transformers}} library \cite{reimers-2019-sentence-bert} to embed both datasets. After that, we calculate the distance between these two embeddings using two widely-adopted metrics as \citet{yueSyntheticTextGeneration2023} does: 
(1) MAUVE\footnote{\href{https://github.com/krishnap25/mauve}{https://github.com/krishnap25/mauve}} \cite{pillutla-etal:mauve:jmlr2023, pillutla-etal:mauve:neurips2021}: MAUVE first clusters the samples in each dataset into a histogram (i.e. two histograms for two datasets), and then uses divergence frontiers \cite{liu-etal:mauve-theory:neurips2021} to calculate the divergence between the two histograms.
(2) Fréchet Inception Distance (FID) \cite{heusel2018gans}: FID calculates the feature-wise mean and covariance matrices of the embedding vectors and then measures the Fréchet distance between the two sets. 

Note that the experiments in Section \ref{sec:data_quality} are based on the same datasets we generated in Section \ref{sec:medical_freeform}. For paraphrase-MiniLM-L6-v2, its FID score is about 10 times the absolute value of other embedding models. Therefore for an unbiased comparison, we scale its score to match the magnitude of others. 

\begin{table*}[t]
\vspace{0mm}
\centering
\small
\setlength\tabcolsep{3pt}
\begin{tabular}{l|cc|cc|cc|cc}
\toprule
\multirow{2}{*}{Baselines} & \multicolumn{2}{c|}{Paraphrase-MiniLM-L6-V2} & \multicolumn{2}{c|}{All-Mpnet-Base-V2} & \multicolumn{2}{c|}{All-MiniLM-L6-V2} & \multicolumn{2}{c}{Avg}\\
& MAUVE ($\uparrow$) & FID ($\downarrow$) & MAUVE ($\uparrow$) & FID ($\downarrow$) & MAUVE ($\uparrow$) & FID ($\downarrow$) & MAUVE ($\uparrow$) & FID ($\downarrow$)\\
\midrule
ICL & 69.83  & 59.96  & 71.73  & 52.33  & 85.00  & 53.76  & 75.52  & 55.35   \\ 
Self-Instruct & 72.26  & 61.27  & 91.72  & 50.05  & 67.72  & 52.82  & 77.07  & 54.21   \\ 
Self-Instruct-ICL & 71.77  & 59.75  & 77.61  & 53.49  & 78.55  & 53.06  & 76.14  & 55.94   \\ 
DP-Gene & 83.23  & 59.41  & 89.58  & 51.42  & 84.47  & 53.58  & 85.76  & 54.80   \\ 
DP-Instruct & 81.29  & \textbf{58.92}  & 83.18  & 50.10  & 89.14  & 51.95  & 84.54  & 53.66   \\ 
DP-Instruct-ICL & 81.97  & 60.00  & 92.20  & \textbf{49.45}  & 82.06  & 52.36  & 85.41  & 53.94   \\ 
KnowledgeSG & \textbf{90.77}  & 59.01  & \textbf{96.48}  & 50.04  & \textbf{92.82}  & \textbf{51.75}  & \textbf{93.36}  & \textbf{53.60}  \\ 
\bottomrule
\end{tabular}
\caption{Embedding distribution distance between the synthetic and original data measured by the MAUVE and FID score. Better similarity indicates better quality of the synthetic data. The results on average reaffirm that \textit{KnowledgeSG} has best data quality compared to other baselines.}
\label{tab:data_embed}
\vspace{-3mm}
\end{table*}

\begin{figure}[t]
\vspace{0mm}
\centering
\setlength{\abovecaptionskip}{0pt} 
\includegraphics[width=\columnwidth]
    {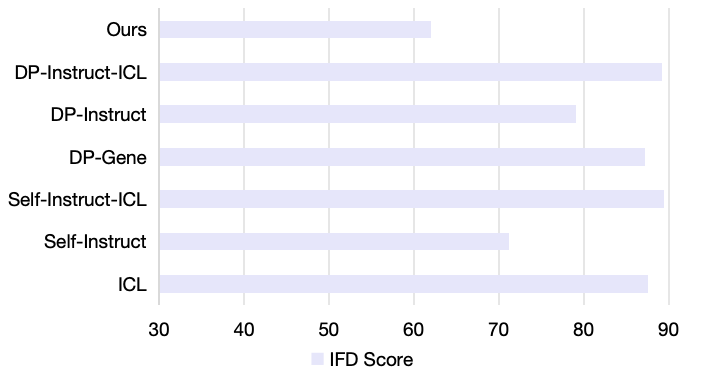}
\vspace{-2mm}
\caption{Instruction following difficulty of different baselines exploiting Llama2-7B as the base model. Lower IFD score indicates better quality of synthetic data. We evaluate on the synthetic datasets which are generated during the experiments in Section \ref{sec:medical_freeform}.}
\label{tab:data_ifd}
\vspace{-3mm}
\end{figure}

\paragraph{Instruction Following Difficulty.}
Instruction following difficulty (IFD) introduced by \cite{Li2023FromQT}, evaluates how much help the instruction provides for the generation of corresponding response. It compares the change of losses in model responses with and without the instructional context, and outputs a ratio as the final score.
A lower IFD score indicates better quality of the evaluated sample. Thus we apply IFD score to measure the utility and quality of the generated instruction tuning datasets. 
The average IFD scores of dataset samples before filtering are presented in Table \ref{tab:data_ifd}, exhibiting the disparity in the generation capabilities across various baselines. 
In practice, we deploy IFD score as the data filtering measure \cite{li2024superfiltering, zhangFedPITPrivacypreservingFewshot2024} in our framework. However, in consideration of fair comparison with other baselines, we exclude it from the experiments in Sections \ref{financial} and \ref{sec:medical_freeform}.

\paragraph{Results.}

From Table \ref{tab:data_embed} and Fig \ref{tab:data_ifd}, We can conclude that:
(1) Although the absolute values of MAUVE and FID are influenced by the specific settings used in its calculation, e.g. scalar scaling constants, the relative rankings of different synthetic datasets remain consistent.
Still, \textit{KnowledgeSG} achieves the best similarity measured by the MAUVE score. For the FID score, our method is only second to DP-Instruct-ICL, an improved version we adopt from \citet{yuPrivacyPreservingInstructionsAligning2024}.
(2) The leading performance of \textit{KnowledgeSG} indicates better quality of synthetic data compared to other baselines. This is consistent with the performance results in Section \ref{sec:medical_freeform}
(3) For instruction following difficulty, the results conform to those of embedding distribution similarity, further proving the effectiveness of our proposed method.

\subsection{Ablation on Dataset Size}
\label{sec:size}
\paragraph{Setups.}
We perform an ablation study on dataset size to investigate its impact on the model's final performance through synthetic data generation. The training and evaluation setups are the same as Section \ref{sec:medical_freeform}. 
For a fair comparison, we make sure that each data sample is iterated $5$ times by training the models for corresponding rounds wile keeping other parameters fixed (e.g., the 500-sample dataset is trained for $50$ rounds, and the 1000-sample dataset for $100$ rounds).

\paragraph{Results.}
For all methods shown in Table \ref{tab:dataset_size}, the results indicate that as the amount of involved data increases, the performance of the trained model improves correspondingly. 
However, the last row of \textit{KnowledgeSG} suggests that the improvement from accumulating additional data may reach a potential threshold. We leave further exploration of this for future work.

\begin{table}[t]
\vspace{0.4mm}
\centering
\small
\begin{tabular}{l|cccc}
\toprule
Dataset Size & 500 & 1000 & 2000 & 3000  \\ 
\midrule
Non-Private & 0.325  & 0.371  & 0.379  & 0.391   \\ 
ICL & 0.329  & 0.335  & 0.364  & 0.368   \\ 
KnowledgeSG & 0.708  & 0.724  & 0.747  & 0.757   \\ 
\bottomrule
\end{tabular}
\caption{Ablations on dataset size. With more data involved, the model performance improves as expected.}
\label{tab:dataset_size}
\vspace{-2.5mm}
\end{table}

\subsection{Transmitting Unit}
\label{sec:transmit_unit}
\paragraph{Setups.}
We employ alpaca \cite{peng2023instruction} and randomly select $50$ samples to form our seed dataset $\mathbb{D}_{Seed}$.
We first fine-tune Llama2-7B on $\mathbb{D}_{Seed}$, then replace the original model with its fine-tuned version.
We assume the attacker only has access to the transmitting process, meaning he can intercept the LoRA adapter fine-tuned on the new base model.
Without access to $\mathbb{D}_{Seed}$, the attacker can only attempt to merge the adapter with the original base model, i.e. open-sourced Llama2-7B, thus unable to reproduce the full performance of our model
\textit{Relative Drop} is calculated by $Relative \ Drop = \frac{(KnowledgeSG - Attacker)}{KnowledgeSG}$.
\paragraph{Results.}
Results in Table \ref{tab:transmit_unit} show that the  performance of model stolen by the attacker drops significantly compared to \textit{KnowledgeSG}. This demonstrates that our model is not compromised, confirming the efficacy of proposed transmitting unit.
\begin{table}[t]
\centering
\small
\begin{tabular}{l|cccc}
\toprule
\multirow{2}{*}{Evaluation} & \multicolumn{2}{c}{Avg:3} & \multicolumn{2}{c}{Avg:4} \\
~& Acc & F1 & Acc & F1 \\
\midrule
Llama2-7B & 0.557  & 0.495  & 0.563  & 0.497   \\ 
KnowledgeSG & 0.784  & 0.775  & 0.752  & 0.745   \\ 
Attacker & 0.419  & 0.343  & 0.428  & 0.350   \\ 
Relative Drop & 46.49\% & 55.76\% & 43.06\%	& 53.08\% \\
\bottomrule
\end{tabular}
\caption{Experiments of proposed transmitting unit. 
The \textit{Relative Drop} in performance suggests that our model is safeguarded against the attacker during transmission.
}
\label{tab:transmit_unit}
\vspace{-2.5mm}
\end{table}

\vspace{-3.8mm}
\section{Discussions}
\subsection{Why not Scrubbing}
The most intuitive way of privacy-preserving is PII scrubbing. 
PII scrubbing is a dataset curation technique that removes PII from text, relying on Named Entity Recognition (NER) to tag PII.
In practice, using scrubbing to mask or add noise to original data, is flawed and must balance the trade-off between minimizing disclosure and preserving the utility of the dataset.
Nonetheless, modern NER has mixed recall of $97\%$ for names and $80\%$ for care unit numbers on medical data \cite{Vakili2022DownstreamTP, lukasAnalyzingLeakagePersonally2023}, indicating that many PIIs are still retained after scrubbing. 

\subsection{Why not DP-SGD only}
Fine-tuning models to satisfies DP can only address the risk of memorization. There is no protection during the data collection stage where the user instructions are exposed to human annotators for response generation \cite{yuPrivacyPreservingInstructionsAligning2024}.
Moreover, using DP-SGD to prevent memorization by adding noise into the training process is destined to sacrifice performance.
As proved in our experiments in Table \ref{tab:dp}, employing DP-SGD alone leads to considerable performance drop.

\section{Conclusions}
This paper addresses the challenge of preserving privacy while fine-tuning large language models on sensitive data. To improve the quality of synthetic data, an aspect often overlooked in previous works, we introduce a novel client-server framework called \textit{KnowledgeSG}. Specifically, \textit{KnowledgeSG} leverages knowledge distillation from a professional server, by prompting it to provide judgments and corrections for raw synthetic data generated by the DP-finetuned base model. Inspired by federated learning, \textit{KnowledgeSG} transmits models rather than data through a specially designed transmitting unit to ensure privacy. 
We conduct extensive experiments, and the results validate the effectiveness of \textit{KnowledgeSG}. The framework achieves a relative improvement of $120.39\%$ compared to the Non-Private training, as measured by medical free-form evaluation. Additionally, \textit{KnowledgeSG} significantly reduces the reconstruction rate from 97.13 to 0.87, demonstrating its strong privacy-preserving capabilities.

\section{Limitations}
While \textit{KnowledgeSG} offers best privacy and performance trade-off across various domain-specific scenarios, its effectiveness on general tasks remains to be fully explored. Further experiments are needed to test its generalizability in broader contexts. 

Also, \textit{KnowledgeSG} involves more communication and computation cost than Non-Private fine-tuning, as it requires DP-finetuning the base model and leveraging a professional model for knowledge distillation. However, we believe these costs are justified, given the significant reduction in memorization concerns and the substantial performance improvements.

For future directions, we plan to conduct experiments on more general tasks and seek ways to optimize communication and computation costs. Additionally, we aim to make the deployment of \textit{KnowledgeSG} more compatible and lightweight.

\section*{Acknowledgments}
This research is supported by the National Key R\&D Program of China under Grant 2021ZD0112801, NSFC under Grant 62171276 and the Science and Technology Commission of Shanghai Municipal under Grant 21511100900 and 22DZ2229005. 
We are grateful to Yifei Zhang, Changyu Miu, Huiyao Chen and Mengying Yuan, for their valuable discussions as well as feedback on the manuscript. We also thank TruthAI, for its GPU support.

\bibliography{privacy_llm, domain}
\vspace{2mm}

\appendix
\section{Privacy Analysis}
\label{app:privacy_analysis}
\subsection{Potential Privacy Risks}
There is a potential privacy concern that the base model may have already encountered the private dataset $\mathbb{D}_{Pri}$ during pre-training. If this is the case, synthetic data generated by the base model $\mathbb{W}_{Loc}$ or its DP-finetuned variant $\mathbb{W}_{DP}$ may still violate privacy requirements \cite{igamberdiev-etal-2022-dp}. Additionally, if the professional model $\mathbb{W}_{Pro}$ has been trained on $\mathbb{D}_{Pri}$, it could inadvertently produce sensitive information such as individual names, when we utilize it to distill knowledge and improve the synthetic data generated by $\mathbb{W}_{DP}$.

To address this concern in \textit{KnowledgeSG}, we will provide both theoretical elaborations and experimental results.
It is important to note that the likelihood of private datasets being leaked and pre-trained by models is minimal in real-world applications. Our work focuses on preventing further memorization when using sensitive data, rather than reversing any memorization that has already occurred.
\subsection{Theoretical Privacy Elaborations}
\paragraph{Interchangeability of Models.}
In our framework, both the base model and professional model are interchangeable. \textit{KnowledgeSG} is not dependent on any specified LLM, e.g. Llama2-7B. The clients using \textit{KnowledgeSG} can select any other LLM that has not been pre-trained on their private datasets to mitigate the risk. 

\paragraph{Theoretical Guarantee of Differential Privacy.}
Based on previous works, we assert the privacy-preserving nature of our framework is justified by differential privacy theory.
First, on the client side, we follow \citet{Abadi_2016, yueSyntheticTextGeneration2023} to DP-fintuned the base model $\mathbb{W}_{Loc}$. This provides us with a strong theoretical guarantee against memorization within the privacy budget $(\epsilon,\delta)-DP$.

Second, on the server side, the post-processing property of DP \cite{Dwork2014TheAF} ensures that once the model $\mathbb{W}_{Loc}$ has been fine-tuned with DP, sampling from the fine-tuned model $\mathbb{W}_{DP}$ does not result in extra privacy loss. Therefore, when the LoRA adapter $\mathbb{A}_{DP}$ is uploaded to the server, it can generate synthetic data without exceeding the privacy budget, mitigating associated privacy risks.

\begin{table}[t]
\centering
\small
\begin{tabular}{l|c}
\toprule
Evaluation & GPT-3.5-turbo \\
\midrule
Llama2-7B   & 12.96  \\
Non-Private & 0.254  \\
ICL         & 0.133  \\
KnowledgeSG & \textbf{0.499}  \\
\bottomrule
\end{tabular}
\caption{Free-form evaluation results using medical-ai-chatbot as the private dataset.}
\label{tab:medical_chatbot}
\vspace{-2.5mm}
\end{table}

\subsection{Experimental Results}
\paragraph{Setups.}
To further validate the effectiveness of \textit{KnowledgeSG} and ensure that no private data has been accessed by either the base model or the professional model, we conducted additional experiments using the ai-medical-chatbot dataset\footnote{\href{https://huggingface.co/datasets/ruslanmv/ai-medical-chatbot}{https://huggingface.co/datasets/ruslanmv/ai-medical-chatbot}}, which was collected and released six months later than Llama2-7B and AlpaCare. We adhere to the experimental setups described in Section \ref{sec:medical_freeform} and also employ Llama2-7B as the base model. 
\paragraph{Results.}
The results presented in Table \ref{tab:medical_chatbot}, reaffirm the effectiveness of \textit{KnowledgeSG}, regardless of whether the models had access to the private dataset. It also shows that \textit{KnowledgeSG} can generalize well across different datasets. Additionally, they demonstrate that \textit{KnowledgeSG} generalizes well across different datasets. 
Llama2 trained on the ai-medical-chatbot dataset yields lower scores compared to its training on HealthCareMagic, indicating that the latter dataset may have higher quality.

Llama2 trained on the ai-medical-chatbot dataset yields lower scores compared to its training on HealthCareMagic, suggesting that the latter dataset may have higher quality.

\section{Additional Techniques}
\label{sec:additional_technique}
\subsection{Filtration with Models}
As mentioned in Section \ref{sec:method}, filtration with model means that we prompt the professional model $\mathbb{W}_{Pro}$ with raw instructions for judgments. Then we filter out subpar instructions based on judgements.

For domain-specific settings such as the medical domain, the judgements are mainly based on whether the tested instructions are related to particular medical knowledge. 
We first prompt AlpaCare using the template written in Figure \ref{fig:prompt_5}, then extract judgements from the model outputs.
In experiments, we also try GPT-3.5-turbo as the domain classifier of instructions and receive acceptable results. 

\subsection{Name Substitution}
\label{sec:name_substitution}
In order to discard the possibility that the pre-trained model has already seen those individual names (e.g. \textit{John, Trump}) in our training datasets $\mathbb{D}_{Pri}$, we ask GPT-4 \cite{openai2023gpt4} to generate hundreds of unique names (e.g. \textit{Anastasija, Melangell}) to substitute the original names. This technique addresses the potential privacy risk discussed in Appendix \ref{app:privacy_analysis} and pave the groundwork for accurate experiments in Section \ref{sec:privacy_evaluation}.

To evaluate the name substitution technique, we follow the experimental setups in Section \ref{sec:privacy_evaluation}, and compare reconstruction rates of different baselines before and after name substitution.
The results in Table \ref{tab:privacy} reveal the effectiveness of our approach. 
Before name substitution, there is no distinguished gap between the different models. After name substitution, as expected, the pre-trained Llama2 exhibits no memorization, while the Non-private approach shows high memorization because of fine-tuning over private data. And the memorization issue is addressed through synthetic text generation.

\begin{table}[t]
\centering
\small
\setlength\tabcolsep{3.3pt}
\begin{tabular}{l|cccc}
\toprule
Reconstruction & Llama2-7B & None-Private & Synthetic  \\
\midrule
Before & 40.23 & 43.73 & 42.57  \\ 
After & 1.89 & 96.23 & 3.77 \\ 
\bottomrule
\end{tabular}
\caption{Reconstruction rate comparison \textit{Before} and \textit{After} name substitution using Flair as the NER extraction tool. The expansion of the gap between Non-Private and Synthetic methods validates our name substitution approach.}
\label{tab:privacy}
\end{table}

\section{Additional Experiments}
\label{sec:add_experiment}
\subsection{Medical Benchmarks}
\label{sec:medical_benchmark}
\paragraph{Setups.}
We evaluate the same models as Section \ref{sec:medical_freeform} on $3$ medical question answering benchmarks including MedQA~\cite{jin2021disease}, PubMedQA~\cite{jin2019pubmedqa}, and MedMCQA~\cite{pmlr-v174-pal22a}.
We follow the code base of LMflow\footnote{\href{https://github.com/OptimalScale/LMFlow}{https://github.com/OptimalScale/LMFlow}} \cite{diao2023lmflow} and use the prompt shown in Figure \ref{fig:prompt_1} to inference answers. 
\begin{table}[t]
\centering
\small
\setlength\tabcolsep{2pt}
\begin{tabular}{l|cccc}
\toprule
Evaluation & PubMedQA & MedQA & MedMCQA & Avg \\
\midrule
Non-Private & 41 & 27.57 & 25.79 & 31.45 \\ 
ICL & 40.9 & 28.75 & 15.31 & 28.32  \\ 
Self-Instruct & 44.4 & 24.27 & 19.85 & 29.51  \\ 
Self-Instruct-ICL & 48.1 & 28.91 & 25.51 & 34.17 \\ 
DP-Gene & 43.2 & 26.08 & 22.53 & 30.60  \\ 
DP-Instruct & 36.8 & 26.24 & 26.46 & 29.83  \\ 
DP-Instruct-ICL & 54.5 & 23.88 & \textbf{27.37} & 35.25 \\ 
KnowledgeSG & \textbf{58.3} & \textbf{30.24} & 26.8 & \textbf{38.45} \\ 
\bottomrule
\end{tabular}
\caption{Performance results on medical domain. Comparative analysis of free-form instruction evaluation.}
\label{tab:medical_benchmark}
\vspace{-2mm}
\end{table}

\paragraph{Results.}
From Table \ref{tab:medical_benchmark}, we can conclude that: 
(1) Compared to free-form evaluation in Section \ref{tab:medical_alpacare}, the results on medical benchmarks are more random. Along with the limit of performance ceiling, the gap between different methods are narrowed especially on MedQA and MedMCQA.
(2) Our method still performs the best on average.
\paragraph{Distinctions in Medical Evaluations.}
Compared to the benchmark results in Table \ref{tab:medical_benchmark}, the gap between different baselines is much more pronounced and noticeable in the free-form evaluation in Table \ref{tab:medical_alpacare}, aligning more closely with expectations. We attribute the reasons as:
(1) For MedQA and MedMCQA, the dataset we use is HealthCareMagic, whose purpose is to provide patients with consultant. This may not correspond with the nature of benchmarks to choose the right answer to a medicine-related question.
(2) Benchmark results involve more randomness, thus improving the performance of inferior competitors to some extent.

\subsection{DP-SGD Performance Evaluation}
We follow the details for DP-finetuning in Appendix \ref{sec:train_details} and evaluate its performance on the financial domain, same as Section \ref{financial}.

From the results in Table \ref{tab:dp}, we can conclude that relying on DP-SGD only results in a considerable decline of performance, necessitating our approach of synthetic data generation with knowledge distillation from server.
\begin{table}[t]
\small
\centering
\setlength\tabcolsep{9pt}
\begin{tabular}{l|cccc}

 \toprule
\multirow{2}{*}{Evaluation} & \multicolumn{2}{c}{Avg:3} & \multicolumn{2}{c}{Avg:4} \\
~& Acc & F1 & Acc & F1 \\
\midrule
Non-Private  & 0.699  & 0.719  & 0.689  & 0.703   \\ 
        DP-SGD  & 0.419  & 0.343  & 0.428  & 0.350   \\ 
        KnowledgeSG  & 0.784  & 0.775  & 0.752  & 0.745   \\
\bottomrule
\end{tabular}
\caption{Comparison of Non-Private approach with DP-SGD. The drop in performance validates the limitations of relying on DP-SGD only.}
\label{tab:dp}
\vspace{-2.5mm}
\end{table}

\subsection{Generalizability in Other Domains}
\paragraph{Setups.}
To evaluate the generalizability of \textit{KnowledgeSG}, we conduct additional experiments in the mathematical and code domains.

For the experimental setup of mathematical domain, we utilize $500$ samples from the lighteval/MATH dataset\footnote{\href{https://huggingface.co/datasets/lighteval/MATH}{https://huggingface.co/datasets/lighteval/MATH}}, employing MAmmoTH-7B \cite{yue2024mammoth} as the professional model and Llama2-7B as the base model. Following \citet{yue2024mammoth}, we evaluate models on the GSM8K dataset \cite{gsm8k} using the public benchmark MAmmoTH\footnote{\href{https://github.com/TIGER-AI-Lab/MAmmoTH}{https://github.com/TIGER-AI-Lab/MAmmoTH}}.
For the code domain, we utilize the PythonCodeInstructions-18k dataset\footnote{\href{https://huggingface.co/datasets/iamtarun/python_code_instructions_18k_alpaca}{https://huggingface.co/datasets/iamtarun/python\_code\_
instructions\_18k\_alpaca}}, employing Llama3-8B-Instruct\footnote{\href{https://huggingface.co/meta-llama/Meta-Llama-3-8B-Instruct}{https://huggingface.co/meta-llama/Meta-Llama-3-8B-Instruct}} as the professional model. We evaluate models on HumanEval dataset \cite{chen2021codex} using the bigcode-evaluation-harness benchmark\footnote{\href{https://github.com/bigcode-project/bigcode-evaluation-harness}{https://github.com/bigcode-project/bigcode-evaluation-harness}} \cite{bigcode-evaluation-harness}.

We compare three representative methods: Non-Private fine-tuning, In-Context Learning (ICL), and a simplified version of \textit{KnowledgeSG} that replaces the synthetic responses in ICL with those generated by the professional model $\mathbb{W}_{Pro}$. 
\paragraph{Results.}
As shown in Table \ref{tab:generalizability}, \textit{KnowledgeSG} outperforms ICL and Non-Private methods. The results confirm the effectiveness of \textit{KnowledgeSG} in the math and code domain, further proving its generalizability. 
However, in the code domain, the performance gap between different methods is less pronounced compared to other domains. We attribute this to the suboptimal coding performance of pre-trained Llama2-7B, which may lack the capacity to generalize effectively on coding tasks. This finding aligns with related studies, where experiments on HumanEval are primarily conducted using the Llama2-13B model or larger variants \cite{luo2023wizardcoderempoweringcodelarge, xu2023wizardlm}.
The reason we prefer financial and medical domain than code and math is that math solving and code writing tasks are not directly related to privacy because there usually is no PIIs in these datasets.

Our preference for the financial and medical domains over the code and math domains in our experiments stems from the fact that datasets involving math solving and code writing are not directly related to privacy concerns, as they typically do not contain personally identifiable information (PII). 

\begin{table}[t]
\centering
\small
\begin{tabular}{l|cc}
\toprule
Evaluation & GSM8K & HumanEval \\
Metric & Accuracy & Pass@10 \\
\midrule
Llama2-7B   & 12.96 & 17.68 \\
Non-Private & 21.30 & 18.90 \\
ICL         & 14.27 & 18.29 \\
KnowledgeSG* & \textbf{33.83} & \textbf{20.73} \\
\bottomrule
\end{tabular}
\caption{Performance results on mathematical and code domains. The relative improvement of KnowledgeSG over Non-Private and ICL demonstrates the generalizability of \textit{KnowledgeSG}. We show accuracy and Pass@10 for GSM8K and HumanEval respectively. \\{*: Given that privacy concerns are not the primary issue in the generation of synthetic data for mathematical and code domains, we adopt a simplified version which focuses on knowledge distillation for convenience. This approach excludes differential privacy fine-tuning, instruction filtration, and the transmitting unit.}}
\label{tab:generalizability}
\vspace{-2mm}
\end{table}

\section{Definition of PII}
\label{sec:pii}
There are various definitions of \textbf{Privacy} catering to different privacy concerns in different scenarios. A LLM can know your preference by digging into your search histories. It can also infer that you have a girlfriend from your recent query of buying flowers on Valentine's day.
In this work, we mainly research on one of the definitions of privacy, i.e. PII which is well-studied by the community.

PII is short for Personal Identifiable Information, representing data that can identify an individual. As detailed elaborated in \citet{lukasAnalyzingLeakagePersonally2023}, PII can be a direct identifier when leakage of that data alone is sufficient to re-identify an individual, or quasi-identifier when only an aggregation of many quasi-identifiers can reliably re-identify an individual. Apart from names and addresses, PII could also be ticker symbol, transaction figures and credit securities accounts in financial domain, and health insurance card numbers in medical domain.

We show examples of PII from HealthCareMagic dataset in Fig \ref{fig:icliniq}. Since our current focus is not on any specific category of leaked PII, we only evaluate Individual Name in Section \ref{experiment} for convenience.

\begin{figure}[t]
\setlength{\abovecaptionskip}{0pt}
\centering
\begin{AIbox}[width=0.48\textwidth]{}
\parbox{1\textwidth}{
\scriptsize
\begin{alltt} 
\textbf{[ Patient's question reveals patient's PII name. ]}\\Patient: "Hi my name is \hl{Anastasija}. I've been having an issue for ..."\\Doctor: "Hello. Thanks for query ..."\\

\textbf{[ Patient's question reveals doctor's PII name. ]}

Patient: "Dear \hl{Dr Eluned}. I would like to ask you..."

Doctor: "Hello and welcome to Chat Doctor ..."\\

\textbf{[ Doctor's answer reveals patient's PII name. ]}

Patient: "Hi, and thanks for checking up on me ..."

Doctor: "Hi \hl{Elaine}, Thanks for asking ...."

\end{alltt}}
\end{AIbox}

\caption{Examples of individual names contained in the ICliniq dataset \cite{li2023chatdoctor}. Individual names as one form of PII, can be used to identify corresponding individuals. For anonymity, we substitute the original names with synthetic ones as mentioned in Appendix \ref{sec:name_substitution}.}
\label{fig:icliniq}
\vspace{-2.5mm}
\end{figure}

\section{Differences of Domain-Specific Data from General Data}
\label{sec:gap}
\begin{table*}[t]
\centering
\small
\begin{tabular}{l|cccccc}
\toprule
~ & Chatbot Arena\footnote{\href{https://chat.lmsys.org/?leaderboard}{Results come from https://chat.lmsys.org/?leaderboard}} & MT-Bench & MMLU & Datum & FPB & PubMedQA  \\ 
\midrule
GPT-4 & 1189 & 8.96 & 86.4 & - & 0.833 & -  \\ 
ChatGPT & - & - & 70.0 & - & 0.781 & 63.9*  \\ 
Llama2-7B-Chat & 1037 & 6.27 & 45.8 & - & - & -  \\ 
Llama2-7B & - & - & - & - & 0.39 & 7.2  \\ 
Llama-7B & - & - & 35.2 & - & - & 5.2*  \\ 
\midrule
Gap Ratio & 0.8722$^\uparrow$ & 0.6998$^\uparrow$ & 0.5301$^\uparrow$ & 0.5$^-$ & 0.4682$^\downarrow$ & 0.1127$^\downarrow$ \\
\bottomrule
\end{tabular}
\caption{Comparison between \{Llama2-7B, Llam2-7B-Chat\} and \{GPT-4, ChatGPT \} on general benchmarks including Chatbot Arena Leaderboard, MT-Bench, MMLU \cite{chiang2024chatbot, mmlu, zheng2023judging} and domain-specific benchmarks including FPB, PubMedQA\cite{Malo2014GoodDO, jin2019pubmedqa}. Results with tagger* is collected from \citet{zhang2023alpacareinstructiontuned}. Results with $^\uparrow$ and $^\downarrow$ indicate whether the \textit{Gap Ratio} exceeds the datum line of $0.5$ or not.}
\label{tab:llama_gpt}
\vspace{-2mm}
\end{table*}
\subsection{Illustration}
We give additional illustration in this section to explain the performance discrepancies of domain-specific data and general data after synthetic data generation.

\begin{figure}[t]
\centering
  \includegraphics[width=1\columnwidth]
  {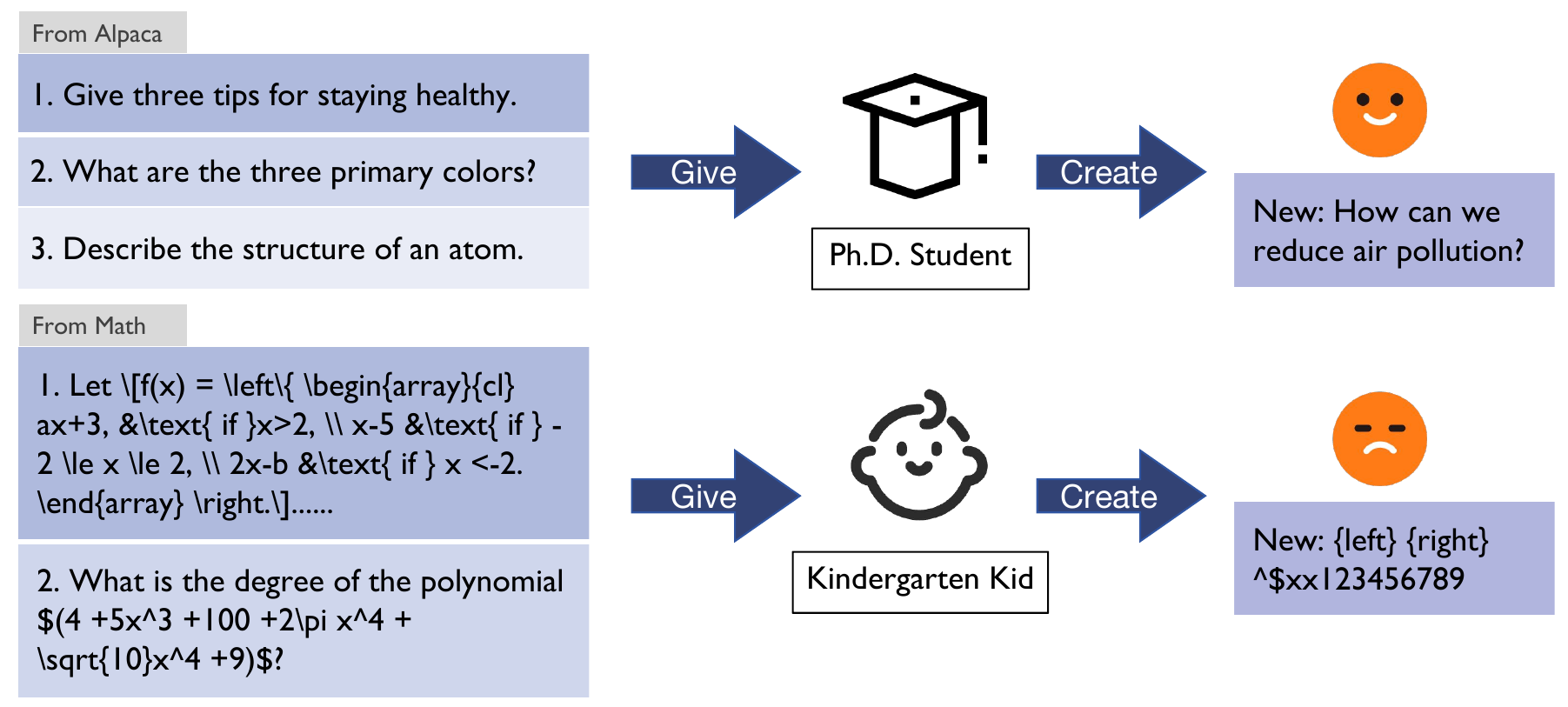}
  
  \caption{Illustration of our identified gap between model comprehension and data complexity. We make an analogy by describing a situation where a student is asked to create a new question based on given examples.}
  \label{fig:ilus}
  \vspace{-2.5mm}
\end{figure}

Deploying an LLM to generate new synthetic data from the original private data is just like asking a student to read an examination question and try to create a new copy of it.
Naturally, the quality of the rewritten question is highly dependent on how the student understands the original question, and how he may generalize.
As illustrated in Fig \ref{fig:ilus}, a Ph.D. student will behave well on general questions, e.g. Alpaca\footnote{\href{https://huggingface.co/datasets/tatsu-lab/alpaca}{https://huggingface.co/datasets/tatsu-lab/alpaca}} \cite{alpaca}.
But if you ask a kindergarten student to create a new calculus test based on several examples, e.g. Math\footnote{\href{https://huggingface.co/datasets/lighteval/MATH}{https://huggingface.co/datasets/lighteval/MATH}} \cite{hendrycksmath2021}, it is highly unlikely that he can fulfil this task.

In practical applications, it is the same nature for LLM-based synthetic data generation where domain-specific data, i.e. the calculus test is more difficult for general foundation models to comprehend. In real-world scenarios when a financial or medical facility tries to train a domain-specific LLM without memorizing its high-value private data \cite{nakamura2020kart, brown2022does}, he is inclined to deploy the synthetic text generation approach. With consideration of resources, he has no choice but to fine-tune a limited-size LLM. However, due to the speciality of original data, small models pre-trained on general data (e.g. Llama2-7B \cite{llama, llama2} and ChatGlm 6B \cite{du2022glm}) are unable to fully understand the domain knowledge and consequently fail to maintain high utility of original data after synthetic generation. 

\subsection{Gap Ratio}
\label{sec:gap_ratio}
For the purpose of quantifying the gap between domain-specific data and general data 
and providing better understanding of the proposed problem, we heuristically define a ratio called \textit{Gap Ratio}. 

We choose GPT-4 \cite{openai2023gpt4} to be the datum model as we assume it is an all-around player that behaves well both on general tasks and domain-specific tasks. And the \textit{Gap Ratio} is calculated by the ratio of target model results and GPT-4 results on the same evaluation benchmark.
For example, from Table \ref{tab:llama_gpt}, Llama2-7B's \textit{Gap Ratio} is $0.8722$ on Chatbot Arena and $0.7007$ on general benchmarks on average. 

No matter what the absolute value is in different measurements of model performance, we can apparently see that the gap between Llama2 and GPT will be greatly widened if changed from general to a specific domain.
As in Table \ref{tab:llama_gpt}, we draw a datum line of $0.5$, smaller than which indicates a tendency of worse synthetic generation.

\section{Implementation Details}
\label{sec:details}
\subsection{Training Details}
\label{sec:train_details}
For normal fine-tuning (not DP), we follow the codebase of \cite{yeOpenFedLLMTrainingLarge2024}\footnote{\href{https://github.com/rui-ye/OpenFedLLM}{https://github.com/rui-ye/OpenFedLLM}} and use the local training algorithm to train the model for $100$ rounds in total.  
For each round, we train for $10$ steps with batch-size set to $5$ using AdamW~\cite{adamw} optimizer. This means each sample in the training dataset is iterated for $10$ times on average, equal to training the model for $10$ epochs without setting max-steps.
We apply a cosine learning rate schedule according to the round index.
The initial learning rate in the first round is $5e-5$, and the final learning rate in the last round is $1e-6$.

For DP fine-tuning, we follow the codebase of dp-transformers library \cite{dp-transformers}\footnote{\href{https://github.com/microsoft/dp-transformers}{https://github.com/microsoft/dp-transformers}}, which is a wrapper around Opacus \cite{opacus}\footnote{\href{https://github.com/pytorch/opacus}{https://github.com/pytorch/opacus}}.
We train the model for $4$ epochs for the first stage of generation, and $10$ epochs for fair comparison between training on private data with DP and training on synthetic data. 
The target epsilon is set to $8$ and maximum per-sample gradient norm is set to $1.0$ for differentially private training. The privacy budget we use is $(\epsilon, \delta) = (8, \frac{1}{N})$. According to \cite{lukasAnalyzingLeakagePersonally2023}, these values are close to established DP deployments such as Apple’s QuickType and Google’s models.

The max sequence length is set to $512$ for training in both normal and DP fine-tuning.
All the training experiments are conducted on one NVIDIA GeForce RTX 3090. 

The rank of LoRA~\cite{hu2021lora} is $32$ with a scalar $\alpha=64$.
We use the Alpaca~\cite{alpaca} template to format the instruction.

\subsection{Inferencing Details}
We use VLLM \cite{kwon2023efficient} for faster inferencing and set the max-model-len to as long as $2048$ to obtain more information. 
The inferencing experiments are mostly conducted on A100 40G.
We set temperature to 0.7 to encourage diversity. 
We follow in-context learning \cite{in-context} and self-instruct \cite{wang2022self} to formulate our prompts. The prompt templates we employ are shown in Figure \ref{fig:prompt_2} and \ref{fig:prompt_3}.
To make sure we have enough instructions for subsequent filtering, the generation times are set two times of the original dataset size. To ensure sufficient instructions for subsequent filtering, the generation count is set to twice the size of the original dataset.
For instruction extraction and pre-processing, we extract the first instruction the model generates and filter those shorter than 2 tokens. 

\subsection{Baselines}
\label{sec:appendix_baseline}
To give a detailed comparison between different baselines in our experiments, we elaborate on three aspects in Table \ref{tab:baseline}, ranging from the model used for generating instructions, whether the baseline first generates instructions then responses and whether the baseline requires few-shot examples to generate response if it is twp-step.
DP-Instruct-ICL and Self-Instruct-ICL are different from DP-Instruct and Self-Instruct in that they require few-shot examples from original dataset to produce better responses during the second stage of generation while the others do not. 
Theoretically, DP-Instruct performs better than Self-Instruct and DP-Gene performs better than ICL because of additional DP-finetuning of base model.
\begin{table}[t]
\centering
\small
\setlength\tabcolsep{4pt}
\begin{tabular}{l|ccc}
\toprule
Baselines & Model & Two-Step & ICL \\
\midrule
ICL & Pre-trained & \XSolidBrush & -  \\ 
Self-Instruct & Pre-trained & \Checkmark & \XSolidBrush  \\ 
Self-Instruct-ICL & Pre-trained & \Checkmark & \Checkmark  \\ 
DP-Gene & DP-finetuned & \XSolidBrush & -  \\ 
DP-Instruct & DP-finetuned & \Checkmark & \XSolidBrush  \\ 
DP-Instruct-ICL & DP-finetuned & \Checkmark & \Checkmark  \\ 
KnowledgeSG & DP-finetuned & \Checkmark & \XSolidBrush  \\
\bottomrule
\end{tabular}
\caption{Elaboration of baselines. \textit{Model} means the generative model used for generating synthetic instructions. \textit{Twp-Step} means whether the baseline first generates instructions then responses or generates both instructions and responses meanwhile. \textit{ICL} means whether the baseline requires few-shot examples from original dataset to generate response at the second stage.}
\label{tab:baseline}
\vspace{-2mm}
\end{table}

\section{Deployment Guidance}
To facilitate real-world applications and future work, we provide a detailed guidance on the deployment of \textit{KnowledgeSG}. The framework involves three main stages.

\paragraph{Preparations and Transmitting Unit.}
(1) Prepare the base model, e.g. Llama2-7B and establish a code base that can do normal-finetuning of LLMs, e.g. LlamaFactory.
(2) Establish a communication channel and sample a small amount of data to construct the seed dataset sharing between the client and server.
(3) Fine-tune the base model on this seed dataset to obtain a modified base model on both client side and server side.

\paragraph{Client Side.}
(1) Prepare the private dataset intended for use.
(2) Establish a code base that can achieve DP-finetuning of LLMs. 

\paragraph{Server Side.}
(1) Prepare the professional model. Most of open-sourced large language models can be easily downloaded from the HuggingFace website.
(2) Write a code that can inference LLMs and design the prompts which are related to the professional model we choose.

After this deployment, we can apply \textit{KnowledgeSG} in a client-server framework and obtain the desired model.

\section{Templates}
\label{sec:template}
\begin{figure}[t]
\setlength{\abovecaptionskip}{0pt}
\parbox{0.5\textwidth}
{\begin{response}
\scriptsize

Below is an instruction that describes a task. Write a response that appropriately completes the request.\\
\\
\#\#\# Instruction:\\
\{Instruction\}\\
\\
\#\#\# Response:

\end{response}}
\caption{Templates-1}
\label{fig:prompt_1}
\vspace{-2mm}
\end{figure}

\begin{figure}[t]
\vspace{-8mm}
\setlength{\abovecaptionskip}{0pt}
\begin{response}
\scriptsize
Based on the following examples, please generate a new and unique example that is different and follows the underlying pattern or theme. Try to make your generation as diverse as possible. \\
\\
\#\# Example:\\
\#\#\# Instruction: \{Instruction 1\} \\
\\
\#\#\# Response: \{Response 1\} \\
\\
\#\# Example:\\
\#\#\# Instruction: \{Instruction 2\} \\
\\
\#\#\# Response: \{Response 2\} \\
\\
\#\# Example:\\

\end{response}
\caption{Templates-2}
\label{fig:prompt_2}
\end{figure}

\begin{figure}[t]
\vspace{-5mm}
\setlength{\abovecaptionskip}{0pt}
\begin{response}
\scriptsize
Come up with a series of tasks:\\

\#\# Example:\\
\#\#\# Instruction: \{Instruction 1\} \\
\\
\#\# Example:\\
\#\#\# Instruction: \{Instruction 2\} \\
\\
\#\# Example:\\
\#\#\# Instruction:
\\
\end{response}
\caption{Templates-3}
\label{fig:prompt_3}
\end{figure}

\begin{figure}[t]
\vspace{-5mm}
\setlength{\abovecaptionskip}{0pt}
\begin{response}
\scriptsize
Come up with examples for the following tasks. Try to generate multiple examples when possible. If the task doesn't require additional input, you can generate the output directly.\\
\\
\{Examples if ICL used\}\\
\\
\#\#\# \{Generated\_Instruction\}\\
\\
\#\#\# Response:
\end{response}
\caption{Templates-4}
\label{fig:prompt_4}
\vspace{-2mm}
\end{figure}

\begin{figure}[t]
\setlength{\abovecaptionskip}{0pt}
\begin{response}
\scriptsize
If you are a doctor, please answer the medical questions based on the patient's description. \\Patient: \{instruction\} Does my instruction invovles medicine?

ChatDoctor: 

\end{response}
\caption{Templates-5}
\label{fig:prompt_5}
\end{figure}
        


        

\end{document}